# The Warped Science of Interstellar

Jean-Pierre Luminet

*Aix-Marseille Université, CNRS, Laboratoire d'Astrophysique de Marseille (LAM) UMR 7326*
*Centre de Physique Théorique de Marseille (CPT) UMR 7332*
*and Observatoire de Paris (LUTH) UMR 8102*
*France*
*E-mail:* `jean-pierre.luminet@lam.fr`

The science fiction film, *Interstellar*, tells the story of a team of astronauts searching a distant galaxy for habitable planets to colonize. *Interstellar*'s story draws heavily from contemporary science. The film makes reference to a range of topics, from established concepts such as fast-spinning black holes, accretion disks, tidal effects, and time dilation, to far more speculative ideas such as wormholes, time travel, additional space dimensions, and the theory of everything. The aim of this article is to decipher some of the scientific notions which support the framework of the movie.

**INTRODUCTION**

The science-fiction movie *Interstellar* (2014) tells the adventures of a group of explorers who use a wormhole to cross intergalactic distances and find potentially habitable exoplanets to colonize. *Interstellar* is a fiction, obeying its own rules of artistic license : the film director Christopher Nolan and the screenwriter, his brother Jonah, did not intended to put on the screens a documentary on astrophysics – they rather wanted to produce a blockbuster, and they succeeded pretty well on this point. However, for the scientific part, they have collaborated with the physicist Kip Thorne, a world-known specialist in general relativity and black hole theory. With such an advisor, the promotion of the movie insisted a lot on the scientific realism of the story, in particular on black hole images calculated by Kip Thorne and the team of visual effects company *Double Negative*. The movie also refers to many aspects of contemporary science, going from well-studied issues such as warped space, fast-spinning black holes, accretion disks, tidal effects or time dilation, to much more speculative ideas which stem beyond the frontiers of our present knowledge, such as wormholes, time travel to the past, extra-space dimensions or the « ultimate equation » of an expected « Theory of Everything ».

It is the reason why, beyond the subjective appreciations that everyone may have about the fiction story itself, many people – physicists and science journalists – have taken the internet to write articles lauding or criticizing the science shown in the movie. Kip Thorne has written a popular book, *The Science of Interstellar*[1], to explain how he tried to respect scientific accuracy, despite the sometimes exotic demands of Christopher and Jonah Nolan, ensuring in particular that the depictions of black holes and relativistic effects were as accurate as possible.

The aim of this article is not to write a (inevitably subjective) review of *Interstellar* as a fiction story, but to decipher some of the scientific notions, which support the framework of the movie.

**AN ARTIFICIAL WORMHOLE IN THE SOLAR SYTEM ?**

In the first part of the film, we are told that a « gravitational anomaly », called a wormhole, has been discovered out near Saturn several decades ago, that a dozen habitable planets have been detected on the « other side » and a dozen astronauts sent to explore them. In particular, one system has three potentially habitable planets, and it is now the mission of the hero, Cooper, to pilot a spaceship through the wormhole and find which planet is more suitable for providing humanity a new home off the dying Earth.

Wormholes arise as mathematical, idealized solutions of the equations of general relativity that describe the shape of space-time generated by a black hole. Their description can be made pictorial by a technique called embedding. As the name suggests, the game is to visualize the shape of a given space by embedding it in a space with an additional dimension. The mapping is useless for the full 4-dimensional space-time of general relativity, as it would have to be embedded in a 5-dimensional space, which is impossible to visualize. But in the case of a non-rotating black hole, the space-time geometry, described by the Schwarzschild metric, remains fixed at all times ; furthermore, due to spherical symmetry, no information is lost by looking only at equatorial slices passing through the center of the sphere. It then becomes easy to visualize all the details of the curvature of the Schwarzschild geometry as a 2-dimensional surface embedded in the usual 3-dimensional Euclidean space. The result comes to a surprise : the embedding surface consists of a paraboloid-shaped throat linking two distinct sheets of space-time. Ludwig Flamm discovered the paraboloid shape as soon as 1916, and Albert Einstein and Nathan Rosen[2] first studied the Schwarzschild throat in 1935. It has a minimum radius equal to the Schwarzschild radius r =2GM/$c^2$, where M is the mass of the black hole, G the gravitational constant and c the speed of light : it corresponds to the event horizon – the immaterial frontier of the black hole – reduced to a circle. For a typical stellar black hole with M = 10 solar masses, we get r = 30 km. It is the size assumed in *Interstellar*.

The Schwarzschild throat (also called the Einstein-Rosen bridge) joins the upper and lower sheets which are perfectly symmetrical, and which we are at liberty to interpret as « parallel universes ». In other words, the throat appears to the upper universe like a black hole consuming matter, light and energy, but for the lower universe it appears as a « white fountain » expelling matter, light and energy.

Now, general relativity determines only the *local* curvature of the space-time, and not its *global* shape. In particular, it allows the two distinct sheets to be different regions of the *same* universe. We have now a black hole and a white fountain in the same space-time at an arbitrary distance from each other, but linked by a stretched-out throat, baptized wormhole by John Wheeler in 1957 (Wheeler also coined the term « black hole », but only ten years later). We can figure out a wormhole like a tunnel with two ends, each in separate points in space-time, and that tunnel would work as a shortcut to travel through space and time. For example, Charles Misner[3] found a solution of the wormhole equations allowing to travel from the solar system to the nearest star, Proxima Centauri, located at 4.2 light-years, in a time shorter than 4.2 years, however without exceeding the speed of light.

Now, coming back to the *Interstellar* story, two major problems arise. First, there is no wormhole without a black hole. Second, is a wormhole traversable, and can it be really used by a spaceship to travel in space-time ?

For the first one, there are clearly several scientific caveats in the scenario. The most common black holes are formed by the gravitational collapse of massive stars, and consequently have a few solar masses. According to the Schwarzschild formula given above, their radius, and consequently that of their putative wormhole, is of order of a few kilometers only, and the tidal forces they generate (see below) are so huge that any spaceship or astronaut would be torn apart at a distance of several thousands kilometers, i.e. well before being able to reach the event horizon and penetrate it.

Besides, since about one star over 10,000 is massive enough to generate a black hole, the latter are pretty rare : it is estimated that the average density of stellar mass black holes in our region of the Milky Way is of the order of 0.00001 per cubic light year. It would thus be extraordinarily improbable that a stellar black hole is so close to us. And even if it were so, at the distance of Saturn, its huge gravitational field would completely destroy the stability of our solar system ! It is the reason why the Nolan's brothers have taken a pure science-fiction option : the wormhole near Saturn has been artificially created by a very advanced civilization, and placed there to help humanity escape the solar system (at the end of the movie, we understand that these aliens were in fact advanced humans from the future ; they created the black hole and its associated wormhole in the first place, manipulating time and events so that things had to unfold the way they did).

The second problem is the traversability of wormholes. The calculations for a non-rotating black hole, described by the Schwarzschild geometry, show that the wormhole is « strangled in the middle » by the infinite gravitational field of a singularity located at the center r=0, so that nothing can pass through it. However, natural black holes, like stars, must be rotating, and it that case the geometry is much more subtle and complex than the Schwarzschild one. It is described by the Kerr metric (as discovered in 1962 by the new Zealand physicist Roy Kerr). And in that case, the mapping technique described above for the Schwarzschild geometry gives a much more complicated and interesting space-time structure, investigated in the 1960's by Brandon Carter (my former PhD advisor) and Roger Penrose[4]. First of all, the central singularity is no more reduced to a central point at r=0, but has the shape of a ring lying in the equatorial plane of the rotating black hole. As a consequence, the ring no more defines an edge of the space-time geometry, because a traveller, apart from the dangers of tidal forces, could come safely within a hair's breadth of the annular singularity provided he does not touch it, or he can even pass through it. The Penrose-Carter diagrams suggest fascinating possibilities for exploration (see e.g. Luminet[5]). Allowable trajectories show that it is theoretically possible to penetrate the interior of a rotating black hole – preferably a very large one, so that the spaceship is not destroyed by tidal forces -, fly above the plane of the singular ring, and escape the black hole by emerging into an exterior universe (which could be topologically the same as the starting one). Along other trajectories, it could also go to the « other side » of the singularity by passing through the ring, and emerge into a completely unknown universe. In such cases, Kerr black holes would have open wormholes, offering fantastic possibilities for space-time travel !

Alas, the Penrose-Carter diagrams are idealized representations of the space-time structure. In the real universe, astrophysical black holes, rotating or not, are formed by physical processes such as gravitational collapse, and in that case, all the general relativistic calculations say that wormholes are unstable : as soon as they are formed, they would also gravitationally collapse and would be not traversable. But this is not quite the end of the story : come into play general relativistic calculations involving also quantum processes. A possible escape has been proposed in 1988 by Thorne and his students[6] : wormholes that are laced with matter or energy exerting an enormous « negative pressure » (*i. e.* a repulsive tension) could be stable and traversable. Such forms of matter or energy are called « exotic ». It happens that, in the framework of quantum mechanics, some energy states of the quantum vacuum obtained by the so-called Casimir effect have the required properties to produce exotic energy. Of course, all of this is highly speculative, but not theoretically impossible – even if the amount of negative energy required maintaining the wormhole open would be greater than the total energy emitted by the Sun during one full year...

Later, other types of traversable wormholes were discovered as allowable solutions to the equations of general relativity[7]. Thus, for a Hollywood sci-fi movie, the screenwriters felt free to imagine that a very advanced civilization could construct a negative wormhole – perhaps by growing a microscopic negative one, as permitted by quantum mechanics –, and use it effectively as a space-time shortcut.

Even allowing this fantastic idea, there is nothing telling us that a spaceship (and the humans inside), made of normal matter, could cross the negative energy region safely. Nevertheless, the theoretical possibility enabled the popular American astronomer-writer Carl Sagan to construct his novel *Contact* (1985) using the idea of communication with extra-terrestrial civilizations by means of wormholes, and already at the time it was Kip Torne who advised Sagan on the possibilities of wormholes. For *Interstellar*, Thorne went beyond and, just to help the team of visual effects, tried to calculate what would give a journey through an artificially created wormhole. Except for the spherical shape of the wormhole, which can be appreciated by the spectators when the spaceship approaches, most of the scientific calculations were not retained by the filmmakers, because of the strange effects generated, arguably non understandable for a general audience. For the rest, we can be sure that, with our present-day knowledge, a traversable wormhole is not only very improbable, but if it existed, its crossing would not at all look like it is shown in the movie !

To conclude this section, it is to be noticed that scientific visualizations of traversable wormholes had already been calculated in 2006 by Alain Riazuelo at the Institut d'Astrophysique de Paris, and the result, available on DVD[8], is much more spectacular than the artistic rendering shown in *Interstellar*.

**THE FAST-SPINNING BLACK HOLE « GARGANTUA »**

Once on the other side of the wormhole, the spaceship and its crew emerge into a three-planets system orbiting around a supermassive black hole called Gargantua. Supermassive black holes, with masses going from one million to several billion solar masses, are suspected to lie in the centers of most of the galaxies. Our Milky Way probably harbors such an object, Sagittarius A*, whose mass is (indirectly) measured as 4 million solar masses (for a review, see Melia[9]). According to Thorne, Gargantua would be rather similar to the still more massive black hole suspected to be located at the center of the Andromeda galaxy, adding up 100 million solar masses[10]. Its size being roughly proportional to its mass, the radius of such a giant would encompass the Earth's orbit around the Sun.

Such enormous black holes are not a science-fiction exaggeration, since we have the observational clues of the existence of « Behemoth » black holes in faraway galaxies. The biggest one yet detected lies in the galaxy NGC 1277, located at 250 million light-years ; its mass could be as large as 17 billion solar masses, and its size would encompass the orbit of Neptune[11].

Another – and very important – characteristic of Gargantua is that it is a fast-spinning black hole. All the objects in the universe – except the universe itself – rotate. Thus a natural black hole must do so, and be described by the Kerr geometry. The latter now depends on two parameters : the black hole mass M and its angular momentum J. An important difference with usual stars, which are in differential rotation, is that the Kerr black holes are rotating with perfect rigidity : all the points on their surface (the event horizon) move with the same angular velocity. There is however a critical angular momentum $J_{max}$ above which the event horizon would « break up » : this limit corresponds to the horizon having a spin velocity equal to the speed of light. For such a black hole, called « extremal », the gravitational field at the event horizon would cancel, because the inward pull of gravity would be compensated by the huge repulsive centrifugal forces.

It is quite possible that most of the black holes formed in the real universe have an angular momentum rather close to this critical limit. For instance, a typical stellar black hole of 3 solar masses, believed to be the engine of many binary X-ray sources, must rotate at almost 5000 revolutions per second. For reasons that will be explained later, the black hole Gargantua shown in *Interstellar* was assumed to have an angular momentum as close as $10^{-10}$ to the critical value $J_{max}$. If theoretically possible, this configuration is physically quite unrealistic, because the more a black hole rotates fast, the more the material orbiting in the same direction is hard to capture, due to the centrifugal forces, while the matter orbiting in the opposite sense is easily sucked into the hole, where it slows the spin. As a consequence, a too-fast spinning black hole would have the tendency to slow down to an equilibrium velocity smaller than that of Gargantua (general relativistic calculations say that black holes spin no faster than about $0.998 J_{max}$).

However, the advantage of a very fast-spinning black hole is that it permits to have planets orbiting extremely close to the event horizon without being swallowed. And this is a key point of the movie, for it also allows for a huge time dilation, see below. For a Schwarzschild black hole (i.e. with angular momentum J=0), the innermost stable circular orbit, inside which any object would spiral and crash into the black hole, is located at 3 times the black hole radius. For a 100 million solar masses black hole, this gives a minimum distance of 900 million kilometers (a little bit more than Jupiter's distance to the Sun). But for a Kerr black hole spinning very close to the critical limit $J_{max}$, the innermost stable circular orbit can be as close as the event horizon itself : 100 million kilometers only. It is the reason why, in *Interstellar*, the closest planet (called Miller) can orbit safely very close to the event horizon without being swallowed.

Another noticeable point is that a Kerr black hole is not a spinning top revolving in a fixed exterior space : as it rotates, it drags the entire fabric of space-time along with it. As a consequence, Miller's planet must orbit at a velocity close to the speed of light.

**ILLUMINATED PLANETS ?**

All right for the gravitational safety of the three-planets system around Gargantua. But where do these planets get heat and light? In principle a star is needed for that, but there is no star around. Heat cannot come from the black hole itself in the form of the Hawking radiation or the recently advanced « firewall » phenomenon[12] : these effects being purely quantum in nature, they could be noticeable for microscopic black holes, but are completely negligible for astrophysical ones. Can light and heat come from the gaseous ring that orbits around Gargantua, called an accretion disk ? Such an explanation is not very consistent with the rest of the story, because later in the movie Cooper inevitably has to go inside the black hole, and he does not get fried. The theory of accretion disks around black holes was actually developed decades ago (for a review, see Abramowicz[13]) and are in agreement with recent astrophysical measurements using gravitational lensing[14]. Because of the incredible forces involved, accretion disks are extremely hot, like millions of degrees hot. They are so brilliant that they can be seen millions of light-years away, and blast out enough radiation to completely destroy any normal material. Thus the astronauts would have been fried as soon as they emerged from the wormhole. Happily for the continuation of the story, it was not the case. So, how the planets can be habitable despite no nearby source of warmth ?

In his popular book as well as in various interviews, Thorne claimed that the light could come from a very « anemic » accretion disk, that has cooled down to the temperature of the Sun basically (5500 °C). Here « anemic » means that the disk has not been fed by new gas (coming for instance from a tidally disrupted star, see below) in the last million years, and that the accretion rate onto the black hole, a critical parameter on which depends the disk luminosity, is extremely low. In that sense, such a quiescent accretion disk could be relatively safe for humans. But I doubt that it could provide enough light and heat to the planet, like the Sun to us, just because an anemic accretion disk would also be optically thin (*i.e.* transparent), while the Sun's photosphere is optically thick (*i.e.* opaque).

**VISUALISATION OF THE ACCRETION DISK**

Since a black hole causes extreme deformations of spacetime, it also creates the strongest possible deflections of light rays passing in its vicinity, and gives rise to spectacular optical illusions, called gravitational lensing. *Interstellar* is the first Hollywood movie to attempt depicting a black hole as it would actually be seen by an observer nearby. For this, the team at Double Negative Visual Effects, in collaboration with Kip Thorne, developed a numerical code to solve the equations of light-ray propagation in the curved spacetime of a Kerr black hole. It allows to describe gravitational lensing of distant stars as viewed by a camera near the event horizon, as well as the images of a gazeous acccretion disk orbiting around the black hole. For the gravitational lensing of background stars, the best simulations ever done are due to Alain Riazuelo[15], at the Institut d'Astrophysique in Paris, who calculated the silhouette of black holes that spin very fast, like Gargantua, in front of a celestial background comprising several thousands of stars.

But perhaps the most striking image of the film *Interstellar* is the one showing a glowing accretion disk which spreads above, below and in front of Gargantua. Accretion disks have been detected in some double-star systems that emit X-ray radiation (with black holes of a few solar masses) and in the centers of numerous galaxies (with black holes whose mass adds up to between one million and several billion solar masses). Due to the lack of spatial resolution (black holes are very far away), no detailed image has yet been taken of an accretion disk ; but the hope of imaging accretion disks around black holes telescopically, using very long baseline interferometry, is nearing reality today via the Event Horizon Telescope[16]. In the meanwhile, we can use the computer to reconstruct how a black hole surrounded by a disk of gas would look. The images must experience extraordinary optical deformations, due to the deflection of light rays produced by the strong curvature of the space-time in the vicinity of the black hole. General relativity allows the calculation of such an effect.

In 1979 I was the first to simulate the black and white appearance of a thin accretion disk gravitationally lensed by a non-spinning black hole, as seen from far away, but close enough to resolve the image[17]. I took a Schwarzschild black hole and a thin disk of gas viewed from the side, either by a distant observer or a photographic plate. In an ordinary situation, meaning in Euclidean space, the curvature is weak. This is the case for the solar system when one observes the planet Saturn surrounded by its magnificent rings, with a viewpoint situated slightly above the plane. Of course, some part of the rings is hidden behind the planet, but one can mentally reconstruct their elliptic outlines quite easily. Around a black hole, everything behaves differently, because of the optical deformations due to the space-time curvature. Strikingly, we can see the top of the disk in its totality, whatever the angle from which we view it may be. The back part of the disk is not hidden by the black hole, since the images that come from it are to some extent enhanced by the curvature, and reach the distant observer. Much more astonishing, one also sees a part of the bottom of the gaseous disk. In fact, the light rays which normally propagate downwards, in a direction opposite to that of the observer, climb back to the top and furnish a « secondary image », a highly deformed picture of the bottom of the disk (in theory, there is a tertiary image which gives an extremely distorted view of the top after the light rays have completed three half-turns, then an image of order 4 which gives a view of the bottom which is even more squashed, and so on to infinity).

Thus, when I saw for the first time the image of the accretion disk in *Interstellar*, I was not surprised to see the disk spreading above, below and in front of Gargantua's silhouette. The visual result was awesome, and the team at *Double Negative* could be proud of that. But when I read in press releases that this image was the first and the more realistic image of a black hole accretion disk ever made, I was puzzled, because basic visual effects were obviously missing.

In my 1979 simulation, I had also taken into account the physical properties of the gaseous disk : rotation, temperature and emissivity. In a thin accretion disk, the intensity of radiation emitted from a given point on the disk depends on its distance from the black hole. Therefore the brightness of the disk cannot be uniform, as suggested in *Interstellar*. The maximum brilliance comes from the inner regions close to the horizon, because it is there that the gas is hottest. In addition, the apparent luminosity of the disk is still very different from its intrinsic luminosity : the radiation picked up at a great distance is frequency- and intensity-shifted with respect to the emitted one. There are two sorts of shift effects. There is the Einstein effect, in which the gravitational field lowers the frequency and decreases the intensity. And there is the better-known Doppler effect, where the displacement of the source with respect to the observer causes amplification as the source approaches and attenuation as the source retreats. In this case, the disk rotating around the black hole causes the Doppler effect. The regions of the disk closest to the black hole rotate at a velocity approaching that of light, so that the Doppler shift is considerable and drastically modifies the image as seen by a faraway observer. The sense of rotation of the disk is such that matter recedes from the observer on, say the right-hand side of the photograph, and approaches on the left-hand side. As the matter recedes, the Doppler deceleration is added to the gravitational deceleration, implying a very strong attenuation on the right-hand side. In contrast, on the left-hand side the two effects tend to cancel each other out, so the image more or less retains its intrinsic intensity. In any case, a realistic image must show a strong asymmetry of the disk's brightness, so that one side is far brighter and the other is far dimmer.

To describe the complete image I obtained (now easily available on the internet[18]), no caption could fit better than these verses by the French poet Gérard de Nerval, written as soon as in 1854:

*In seeking the eye of God, I saw nought but an orbit*
*Vast, black, and bottomless, from which the night which there lives*
*Shines on the world and continually thickens*

*A strange rainbow surrounds this somber well,*
*Threshold of the ancient chaos whose offspring is shadow,*
*A spiral engulfing Worlds and Days !*[19]

All the above-mentioned effects change also the colors, e.g. from blue on one side to red on the other, and so on. This could not be seen on my black and white (bolometric) image, but my pioneering work motivated visualizations of accretion disks around black holes with ever increasing sophistication. Especially, Fukue and Yokoyama[20] added colours to the disk ; Viergutz[21] made the black hole spin and produced coloured images including the disk's secondary image wrapping under the black hole; Marck[22] laid the foundations for a nice movie available on the web[23], with the camera moving around close to the disk, and included higher order images. Sophisticated ray-tracing codes and accretion flow models have been developed recently, including full 3D-simulations of accretion flows and images of these, see for instance Chan et al.[24] for simulating the aspect of the Galactic black hole Sagittarius A*.

Of course Kip Thorne did not ignore these effects. But, as he wrote me in a spontaneous mail, the film director estimated that a general audience would have been totally baffled by what they are looking at ; so a conscious decision was made to leave out the Einstein and Doppler shifts as well as the physical properties of the disk, and have an accretion disk with the right shape but not the right lopsidedness. As an additional simplification, they have also chosen to apply their calculations to a black hole smaller than Gargantua, and with a much more moderate spin – otherwise the visual effects would have become completely weird and incomprehensible, even for educated physicists ! However, in order to fully exploit their ray-tracing code, in parallel to the movie and the popular nook Thorne and the team at *Double Negative* have submitted to a peer-review a technical paper[25] including all the corrections.

**TIDAL STRESS**

When an object – a planet, a star – moves around a black hole, the forces of gravity act more strongly on the side of the body nearer to the black hole than on the other side. The difference between the two forces is called the tidal force. If the celestial body moves along an approximately circular orbit at a reasonable distance, the tidal forces remain small and the body is able to adjust its internal configuration to the external forces, adopting an elongated shape oriented towards the hole. However, if the body is moving along an eccentric orbit, as the distance $r$ from the black hole decreases the tidal forces increase rapidly (like $r^{-3}$). Eventually there comes a point where these forces are as large as the forces binding the body together. The planet or the star have no longer time to adjust their internal configuration, begin to deform catastrophically and are inevitably disrupted. This happens rather frequently in the universe. In the 1980's, I worked a lot on the process of disruption of full stars by massive black holes[26]. For extreme cases, when the star grazes the event horizon along a parabolic orbit without being swallowed, I predicted the occurrence of « flambéed stellar pancakes », releasing a lot of radiative energy[27]. Our telescopes have since captured such scenes. However, these events occur only when the body gets within some critical radius from the black hole, called the Roche limit – after the French mathematician who studied the problem of tidal forces in 1847 in the context of planets and their satellites.

On the Internet, some bloggers have claimed that the Miller's planet should be completely destroyed by tidal forces, since it is so close to the black hole's surface. This is not necessarily so, and *Interstellar* is marginally correct on this point. In effect, the Roche limit depends on the mass of the black hole and on the average density of the external body according to the law $R_R \sim (M/\rho_*)^{1/3}$, where M is the mass of the black hole and $\rho_*$ the density of the body. Applying this formula to the case of Gargantua (M = $10^8$ solar masses) and a water planet ($\rho_* \sim 1$ g/cm$^3$) we get $R_R \sim 10^{13}$ cm. Now the gravitational radius of the Gargantua black hole, $GM/c^2$, is also of order $10^{13}$ cm. Therefore Miller's planet must suffer large tidal forces, but not enough to be torn apart (for black holes still more massive than $10^8$ solar masses, such as those suspected to lie in the centers of quasars, the Roche limit becomes significantly smaller than the gravitational radius, and in that case, planets or stars can be broken up by the tidal forces only once they are inside the black hole).

Now, in the movie, once the explorers have « landed » on Miller's planet - a water one -, they find it suffering periodic and enormous tidal waves sweeping around. These are unexplained, but we can assume they are caused by the tides from the black hole. We would have to solve some equations to find out whether such kilometer high waves are physically realistic. These equations both involve laws of gravity (if their origin is indeed tidal) and hydrodynamics, more precisely the Navier-Stokes theory. From the film, we notice that the wavelength of the water waves is much greater than the depth of the water itself ; such a situation is ripe for the « shallow-water » approximation, which are obtained by applying the Navier-Stokes equations to such a problem. Fluid mechanics textbooks discuss these equations at length ; they are coupled, nonlinear partial differential equations, depending on various parameters such as the surface gravity on the planet *g*, its rotation rate, the viscous drag forces, and so on.

For the situation in *Interstellar*, we are told that the acceleration due to gravity on the planet is 130% that of Earth's, which means that *g* = 9.81 x 1.30 = 12.75 m/s$^2$. The other parameters will be influenced by internal forces in planet's structure, combined with complex external effects due to the gravitational field of the rotating black hole. There are too many uncertainties on the knowledge of all the parameters to be able to solve numerically the equations.

Nevertheless, I suspect some inconsistency in the film. A tidal wave is actually a bulge of water fixed in space, always oriented in the same configuration, so the astronauts on the planet rotate in and out of that bulge. They feel it as a wave coming towards them and away from them, experiencing from a high tide part of the water to a low tide part of the water. In *Interstellar*, the waves come every hour or so, which means that the planet rotates once ever two of those (because there are two high tides for every rotation). The problem is that with such huge tides, the planet should become quickly tidally locked with the black hole (*i.e.*, like the Moon to the Earth, always presenting the same face to the black hole). We have formulas telling how fast tidal locking occurs in binary systems. Using a $10^8$ solar masses black hole and a planet with a surface gravity about 13 m/s$^2$, we find that the time scale for tidal locking is only 1 millisecond ! Once the planet is tidally locked to the black hole, it spins only once per revolution, and on top of it water stays in place, always pulled towards the black hole.

**A HUGE TIME DILATION**

The elasticity of time is a major consequence of relativity theory, according to which time runs differently for two observers with a relative acceleration – or, from the Equivalence Principle, moving in gravitational fields of different intensities. This well-known phenomenon, checked experimentally to high accuracy, is called « time dilation ». Thus, close to the event horizon of a black hole, where the gravitational field is huge, time dilation is also huge, because the clocks will be strongly slowed down compared to farther clocks. This is one of the most stunning elements of the scenario of *Interstellar* : on the water planet so close to Gargantua, it is claimed that 1 hour in the planet's reference frame corresponds to 7 years in an observer's reference frame far from the black hole (for instance on Earth). This corresponds to a time dilation factor of 60,000. Although the time dilation tends to infinity when a clock tends to the event horizon (this is precisely why no signal can leave it to reach any external observer), at first sight a time dilation as large as 60,000 seems impossible for a planet orbiting the black hole on a stable orbit. As explained by Thorne in his popular book, such a large time dilation was a « non-negotiable » request of the film director, for the needs of the story. Intuitively, even an expert in general relativity would estimate impossible to reconcile an enormous time differential with a planet skimming up the event horizon and safely enduring the correspondingly enormous gravitational forces. However Thorne did a few hours of calculations and came to the conclusion that in fact it was marginally possible (although very unlikely). The key point is the black hole's spin. A rotating black hole, described by the Kerr metric, behaves rather differently from a static one, described by the Schwarzschild metric. The time dilation equation derived from the Kerr metric takes the form:

$$1 - (d\tau/dt)^2 = 2GMr/c^2\rho^2, \text{ where } \rho^2 = r^2 + (J/Mc)^2\cos^2\theta.$$

Substituting for $d\tau = 1$ hour and $dt = 7$ years, one obtains the following relation:

$$\frac{1.334 \times 10^{-10} M^3 r}{8.98755 \times 10^{16} M^2 r^2 + J^2 \cos^2\theta} = \frac{3369802499}{3369802500}$$

This equation fully describes a black hole of mass M, rotating with angular momentum J, as observed by an observer at radial coordinate r and angular coordinate θ. The fraction on the right-hand-side fully depicts the 1 hour = 7 years dilation effect. For the Schwarzschild metric, the orbital radius should be no smaller than 3 times the gravitational radius, and such a time dilation could not be achieved for the planet of the film. But as already said, the Kerr metric allows for stable orbits much closer to the event horizon. Calculations indicate that for M = $10^8$ solar masses, we get r = 1.48x$10^{13}$ cm, θ = π and J = 8.80275x$10^{57}$ J.s. This implies a black hole angular momentum J extraordinarily close (at $10^{-10}$) to the maximal possible value $J_{max}$, a circular orbit lying in the equatorial plane and a radius orbit practically equal to the black hole's gravitational radius. All this is theoretically possible, but by no ways realistic.

**A CLEVER USE OF THE PENROSE PROCESS**

Another effect specific to the physics of rotating black holes, which was correctly depicted in *Interstellar*, is the Penrose process. The astronauts use it to benefit of a particularly efficient gravitational assistance (called « slingshot effect »), which allows their spaceship to plunge very close to the event horizon and escape with an increased energy. In effect, the laws of Kerr black hole physics say that, although a black hole prevents any radiation or matter from escaping, it can give up a part of its rotational energy to the external medium. The key role is played by the *ergosphere*, a region between the event horizon and the static limit below which, like in a maelstrom, space-time itself is irresistibly dragged along with it (the so-called « Lense-Thirring effect »). In a thought experiment, Roger Penrose suggested in 1969 the following mechanism[28]. A projectile disintegrates into the ergosphere, one of the fragments falls into the event horizon in a direction opposite to the black hole's rotation, while the other fragment can leave and be recovered, carrying more energy than the initial projectile. Replace the projectiles by a spaceship which leaves a part of it to fall into the black hole along a carefully chosen retrograde orbit, and *le tour est joué*. The calculations indicate that one can extract an energy equivalent to the rest-mass energy of the part lost into the black hole, which, according to the famous formula $E=mc^2$ can already be huge, plus an additional energy extracted from the spinning black hole, which has been slowed down by the infalling fragment in a retrograde orbit. For a black hole like Gargantua, rotating at almost the maximal speed, repeated Penrose processes could extract 29% of its mass.

**TIME TRAVEL INSIDE GARGANTUA**

In the last part of the film, the main character, Cooper, plunges into Gargantua. There, beware the tidal forces breaking anything up ! Indeed in the Schwarzschild geometry, the tidal forces become infinite as r -> 0 ; so, even for a supermassive black hole like Gargantua, once past safely the event horizon and approaching the central singularity, everything will be ultimately destroyed. Happily for the continuation of the story, Gargantua has a high spin, and its lethal singularity has the shape of an avoidable ring. Thus the space-time structure allows Cooper to use the Kerr black hole as a wormhole ; he avoids the ring singularity and transports to another region of space-time. In the movie he ends up in a five-dimensional universe, in which he will be able to go backwards in time and communicate with his daughter by means of gravitational signals.

A lot of research has been done on whether the laws of physics permit travel back in time or not. Black hole physics gives interesting results but no firm answers. As seen above, according to Penrose-Carter diagrams a rotating black hole could connect myriads of wormholes to different parts of the space-time geometry. Since two events can differ in time as well as in space, it would be possible to pass from one given position at a given time, along a carefully chosen trajectory, through a wormhole, and arrive at the same position but at a different time, in the past or future. In other words, the black hole could be a sort of time travel machine.

Noneless a journey back through time is an affront to common sense. It is difficult to accept that a man could travel back through time and kill his grandfather before he has had the time to produce children. For the murderer could not have been born, and could not have murdered him, and so on... Such time paradoxes have been pleasantly presented in the celebrated series of movies *Back to the future*.

A journey into the past violates the law of causality, which requires that the cause always precedes the effect. However, causality is a rule imposed by logic, not by the theory of relativity. Causality is implicit in Special Relativity, where there is no gravitation. Here, travelling into the past requires motion faster than light, and is absolutely forbidden. However in general relativity, the universe is curved by gravitation, and the space-time geometry can be distorted – by a wormhole associated to a rotating black hole, for example –, enabling the past to be explored without having to go faster than light.

Do such strange time warps exist in the Universe ? Perhaps yes at the quantum scale – *i.e.* at a size much less than proton's radius –, due to the quantum fluctuations of space-time. If microscopic black holes were created soon after the Big Bang[29], the laws of quantum physics would govern them, microscopic wormholes would also be created, and some elementary particles could furtively take these transitory tunnels in order to move back in time. It is not theoretically impossible that very special conditions in the early universe lead to the formation of mini-wormholes that have grown to become macroscopic after a phase of primordial inflation. As a consequence, some supermassive black holes, such as the Gargantua depicted in *Interstellar*, could be fossil giant wormholes, namely open gates leading to other regions of the cosmos, or to parallel universes. And why not a 5-dimensional universe ? This is precisely what is assumed in the movie.

**THE FIFTH DIMENSION**

From this moment, the story becomes more than ever speculative – as much as the new theories that subtend it. The standard theory of gravity – general relativity – describes our universe as geometry of three-dimensional space plus one dimension of time. This is sometimes called 3 + 1 space, and it gives a very accurate description of the universe we observe. But theorists like to play around with alternative models to see how they differ from regular general relativity. They may look at 2 + 1 space, a kind of « flatland » with time. There is not necessarily anything « real » about these models, and there is not any experimental evidence to support anything other than 3 + 1 gravity. But alternative models are useful because they help us gain a deeper understanding of gravity.

One of these alternative models stems from the so-called brane cosmologies[30]. The central idea of brane cosmologies is that our 4-dimensional universe is restricted to a « brane » inside a higher-dimensional space, called the « bulk » (an analogue of the science-fiction notion of « hyperspace »). In the brane models, some of the extra-dimensions are possibly infinite, and other branes can be moving through the bulk. Gravitational interactions with the bulk can influence our brane, and thus introduce effects not seen in standard cosmological models.

Lisa Randall and Raman Sundrum[31] have proposed one of these brane cosmologies at the end of the 1990's. There are two different versions of it, but both assume that our 4-dimensional universe is a brane inside a 5-dimensional space-time, the bulk. In such a framework, we can imagine (although very unrealistically) to create an artificial mini-black hole and mini-wormhole, for instance in a powerful particle accelerator such as the CERN's Large Hadron Collider, and make it growing. In a Randall-Sundrum universe, matter and light cannot propagate in the fifth dimension, and gravitational waves are the only physical entities that can propagate in the bulk. It is exactly what is suggested in the movie. The screenwriters have imagined a very advanced civilization born into the bulk, able to master the laws of gravity to create wormholes and influence our usual brane by means of gravitational waves. Since we learn at the end that this advanced civilization is nothing else than our future humanity, we realize that one of the « philosophical » issues of the film is that humanity should tend to understand the laws of quantum gravity and master the new physical effects involved in order to save itself !

**THE FINAL EQUATION**

At the very end of the film, the scientist's character called Murph begins to write an equation aimed to solve the problem of the incompatibility between general relativity and quantum mechanics. We can see blackboards covered by diagrams and equations supposed to be a possible way to the « ultimate equation » of a so-called « Theory Of Everything ». If discovered by the scientists, it would eventually help to solve all the problems of humanity. I will not discuss the naivety of such a view, but briefly discuss the question whether the equations on the screen have any meaning.

At first sight we can doubt because the unification of general relativity and quantum mechanics remains unsolved – even if various approaches, such as the loop quantum gravity[32], the string theory[33] (of which the Randall-Sundrum model referred above is a very particular solution) or the non-commutative geometry[34], are intensively explored by theoretical physicists all around the world.

Clearly the *Interstellar* filmmakers have bet on the most « fashionable » attempt for unifying all fundamental interactions : string theory. String theory stipulates that the fundamental constituents of matter are not point-like particles but open or closed strings on the scale of the Planck length ($10^{-33}$ cm), whose vibrational modes define particle properties. In this framework, space-time becomes a derived concept, which only makes sense at a scale larger than that of the strings. String theory, which comes in five different varieties, aroused such keen interest that in the 1990's, certain theorists believed it was capable of giving a « Theory Of Everything ». However, the mathematical difficulties involved are formidable, and it is not certain that they will be resolved in the future[35]. The five different string theories have given birth to a larger supposed theory, of which the string theories would only be limits : not only one-dimensional lines could vibrate, but also two-dimensional surfaces and other spaces of higher dimension, such as membranes - from whence the name M-theory given to this hypothesis.

Returning to the ultimate equation briefly seen in *Interstellar* on Murph's blackboard, something looking like $S = \int \sqrt{-g_5}\, d^5x \{\mathcal{L}_{bulk} + \cdots\}$ as far as I can remember, physicists who know a little bit of string theory will recognize the so-called effective action of M-theory in the lowest approximation of its perturbative development. In simpler terms, it gives a hint of what would like the « ultimate equation » of physics if M-theory was the correct framework. It is probably not, but I imagine that Kip Thorne made a *clin d'œil* to string theorists to mean that, in his opinion, a future Theory of Everything will probably be similar to M-theory. I do not share this point of view, but it is rather unexpected to find such a sophisticated message in a Hollywood movie.

The $g_5$ and $d^5x$ terms mean that we deal with a theory in 5 dimensions : 1 for time and 4 for space, like in the Randall-Sundrum case. In a 4-dimensional space without curvature, the analog of the cube is called an hypercube or a tesseract. The tesseract is to the cube as the cube is to the square. Just as the surface of the cube consists of 6 square faces, the hypersurface of the tesseract consists of 8 cubical cells, giving a visual effect which has been excitingly represented in *Interstellar*. But for my part, from an artistic point of view, I am much more moved by the *Corpus Hypercubus* painted in 1954 by Salvador Dali. The Spanish artist depicted the cross of crucufixion as a tesseract to signify that, just as God could exist in a space which is incomprehensible to the humans, the hypercube exists in four space dimensions which are equally inaccessible to ordinary minds.

**CONCLUSION**

To summarize, the movie *Interstellar* is appealing by the fact that it tries to combine a great story (saving humankind by interstellar travel) with accurate science, more or less realistic depictions of general relativistic phenomena, and hazardous extrapolations about new physical laws that could stem from quantum gravity scenarios. But we must keep in mind that *Interstellar* is primarily a science-fiction movie, so that artistic license and scientific extrapolation are integrant part of the game. The main interest of discussing its science accuracy is thus educational. Hollywood becoming aware of science and trying to present its ideas in a correct way is good news. A few months earlier one could see *Gravity* and its impressive show of a hostile and weightlessness space. But most of the science shown in *Gravity* could be understood in the framework of Newtonian theory, published more than 400 years ago and assumed to be well-known of everybody. On the contrary, most of the phenomena depicted in *Interstellar* require to understand the basics of General Relativity – the theory gravitation due to Albert Einstein and whose centenarian we shall celebrate in 2015 –, as well as Quantum Mechanics and even a little bit of String Theory … By inviting the spectators to wonder about deep questions on time, space, gravity and so on, *Interstellar* can drag the youngest to consider careers in science rather than in finance or law. It is completely up to a genre called in French « le merveilleux scientifique », that is the adventure of a science pushed to the marvel, or of a marvel envisaged scientifically.

---

[1] K. Thorne : *The Science of Interstellar*, Norton & Company (november 2014).

[2] A. Einstein, N. Rosen :*The Particle Problem in the General Theory of Relativity*, Physical Review **48**: 73 (1935).

[3] C. W. Misner : *Wormhole Initial Conditions*, Physical Review **118**, 1110 (1960).

[4] B. and C. DeWitt (eds) : *Black Holes (Les Houches School 1972)*, Gordon Breach, New York (1973)

[5] J.-P. Luminet : *Black Holes*, Cambrige University Press, chap. 12. For the French readers, the updated version is J.-P. Luminet : *Le destin de l'univers, trous noirs et énergie sombre*, Paris : Fayard (2006).

[6] M. Morris, M. S., Thorne, K. S., and Yurtsever, U. : *Wormholes, Time Machines and the Weak Energy Condition*, Phys. Rev. Letters, **61**, 1446-1449 (1988).

[7] M. Visser : *Lorentzian Wormholes*, American Institute of Physics, 1996.

[8] A. Riazuelo : *Voyage au cœur d'un trou noir*, DVD (Paris : Sciences et Avenir), 2008.

[9] F. Melia : *The Galactic Supermassive Black Hole*. Princeton: Princeton University Press (2007).

[10] R. Bender al. : *HST STIS Spectroscopy of the Triple Nucleus of M31: Two Nested Disks in Keplerian Rotation around a Supermassive Black Hole*. Astrophysical Journal **631** (1): 280–300 (2005).

[11] R. van den Bosch et al. : *An over-massive black hole in the compact lenticular galaxy NGC 1277*, Nature **491**, 729 (2012).

[12] A. Almheiri, D. Marolf, J. Polchinski and J. Sully : *Black holes: complementarity or firewalls?* Journal of High Energy Physics **2013** (2).

[13] M. A. Abramowicz and P. Chris Fragile : *Foundations of Black Hole Accretion Disk Theory*, Living Rev. Relativity, 16 (2013)

[14] S. Poindexter et al. : *The Spatial Structure of An Accretion Disk*, The Astrophysical Journal, **673** (1): 34 (2008).

[15] A. Riazuelo, http://www2.iap.fr/users/riazuelo/interstellar

[16] Event Horizon Telescope Collaboration website: http://eventhorizontelescope.org

[17] J-P. Luminet : *Image of a spherical black hole with thin accretion disk*, Astron. Astrophys. **75,** 228 (1979).

[18] J.-P. Luminet : http://luth2.obspm.fr/~luminet/Books/Nerval.html

[19] G. de Nerval, *Le Christ aux Oliviers*, in *Les Chimères*, Paris, 1854. Free translation by J.-P. Luminet.

[20] J. Fukue and T. Yokoyama : *Colour photographs of an accretion disk around a black hole*, Publ. Astron. Soc. Japan **40** 15 (1988)

[21] S. U. Viergutz : *Image generation in Kerr geometry I. Analytical investigations on the stationary emitter-observer problem,* Astron. Astrophys. **272,** 355 (1993).

[22] J.-A. Marck :*Short-cut method of solution of geodesic equations for Schwarzschild black hole,* Class. Quantum Grav. **13** 393402 (1996). See also Marck J.-A. and Luminet J.-P. : *Plongeon dans un trou noir*, Pour la Science Hors-Série « Les trous noirs » (July 1997) 50-56.

[23] J.-A. Marck : https://www.youtube.com/watch?v=5Oqop50ltrM. Conversion into a movie first appeared in the documentary *Infinitely Curved* by L. Delesalle, M. Lachièze-Rey and J.-P. Luminet, CNRS/Arte, France, 1994.

[24] C.K. Chan et al. (2014) : *The Power of Imaging : Constraining the Plasma Properties of GRMHD Simulations Using EHT Observations of SgrA\** [arXiv :1410.3492].

[25] O. James, E. von Tunzelmann, P. Franklin, K. Thorne : *Gravitational Lensing by Spinning Black Holes in Astrophysics and in the Movie Interstellar*, submitted to Classical and Quantum Gravity (2014)

[26] J.-P. Luminet & B.Carter : *Dynamics of an Affine Star Model in a Black Hole Tidal Field,* Astrophys. J. Suppl. **61**, 219-248 (1986)

[27] B.Carter & J.-P. Luminet : *Pancake Detonation of Stars by Black Holes in Galactic Nuclei,* Nature **296**, 211 (1982)

[28] Penrose, R. : *Gravitational Collapse: the Role of General Relativity*, Rivista del Nuovo Cimento, Numero Speziale 1, 252 (1969).

[29] B.J. Carr, S.W. Hawking : *Black holes in the Eearly universe*, Mon. Not. Roy. Astron. Soc. **168**, 399 (1974).

[30] P. Brax, C. ven de Bruck : *Cosmology and Brane Worlds: A Review,* arXiv:hep-th/0303095 (2003).

[31] L. Randall, R. Sundrum : *Large Mass Hierarchy from a Small Extra Dimension*, Physical Review Letters **83** (17), 3370–3373 (1999).

[32] C. Rovelli : *Loop Quantum Gravity*, Living Reviews in Relativity **1** (1998).

[33] M. Green, J. H. Schwarz, E. Witten : *Superstring theory*, Cambridge University Press (1987).

[34] A. Connes : *Non-commutative geometry*, Boston : Academic Press (1994).

[35] P. Woit : *Not Even Wrong: The Failure of String Theory And the Search for Unity in Physical Law*. New York: Basic Books (2006).